\documentclass[aps,twocolumn,epsfig,graphics,showpacs,floatfix,mathbbm,showpacs]{revtex4}
\usepackage[colorlinks=true, pdfstartview=FitV, linkcolor=red, citecolor=blue, urlcolor=blue]{hyperref}

\usepackage{amsmath,amsfonts,amssymb,graphics,graphicx,epsfig,color,times,bbm}
\usepackage{graphicx}
\usepackage{epstopdf}

\newcommand{\ket}[1]{| #1 \rangle}
\newcommand{\bra}[1]{\langle #1 |}
\newcommand{\braket}[2]{\langle #1 | #2 \rangle}

\newcommand{\id}{\mathbbm{1}}
\newcommand{\heff}{\tilde H}
\newcommand{\ueff}{\tilde U}

\begin{document}

\title{Optical generation of matter qubit graph states} 

\author{S.C.\ Benjamin$^1$,  J.\ Eisert$^{2,3}$, and T.M.\ Stace$^4$,}

\affiliation{1 Materials Department, 
University of Oxford, Oxford OX1 3PH, UK\\
2 Quantum Optics and Laser Science, Blackett Laboratory, Imperial College London,
London SW7 2BW, UK\\
3 Institute for Mathematical Sciences, Imperial College London,
London SW7 2PE, UK\\
4 Department of Applied Mathematics and Theoretical Physics, University of 
Cambridge, Cambridge CB3 0WA, UK}

\begin{abstract}
We present a scheme for rapidly entangling matter qubits in order to 
create graph states for one-way quantum computing.  
The qubits can be simple
$3$-level systems in separate cavities. 
Coupling involves only local 
fields and a static (unswitched)
linear optics network. Fusion of graph state sections occurs with, 
in principle, zero probability
of damaging the nascent graph state. 
We avoid the finite thresholds of other schemes 
by operating on two entangled pairs, so that each generates exactly one 
photon. We do not require the relatively slow single qubit
local flips to be applied during the growth phase: growth of the graph 
state can then become a purely optical 
process. The scheme naturally 
generates graph states with vertices of high degree and so is easily 
able to construct minimal graph states, with consequent resource 
savings. The most efficient approach will be to
create new graph state edges even as qubits elsewhere are measured, in a 
`just in time' 
approach. An error analysis indicates 
that the scheme is relatively robust 
against imperfections in the apparatus.
\end{abstract}


\maketitle

\section{Introduction}
Despite significant and 
exciting experimental 
progress in recent years, the physical 
realization of a full-scale
quantum computer (QC) remains a tremedous challenge 
\cite{Nielsen}.
In many systems excellent single qubits have already been 
realized (notably, of ions in a trap \cite{Blatt,Wineland}, 
NV centres in diamond \cite{NVDiamond,NVDiamond2}, 
etc). However,  few systems have demonstrated controlled 
qubit-qubit coupling between pairs taken from more than 
four qubits, and achieving the necessary 
exquisite control remains highly problematic.
In general it is difficult to simultaneously satisfy the two key 
requirements of
coupling diffent subsystems in a controlled 
manner, while at the same time shielding the 
system from its environment \cite{Nielsen}. 
In the majority of QC schemes, 
some direct physical interaction is supposed to 
generate the two-qubit 
operations (e.g., phonon modes among trapped ions 
\cite{Blatt,Wineland}, 
F{\"o}rster interactions 
between excitions in semiconduction 
quantum dots \cite{Forster}, 
etc). Thus one calls 
for the qubits to strongly interact with selective parts of their 
environment (namely, other qubits and the control mechanisms) while 
avoiding interactions the rest of the environment to a near perfect 
degree. This is obviously an challenging prescription.

As an alternative to employing a direct physical interaction between 
qubits, one can exploit the entangling power of \emph{measurements}.
A suitable measurement, at least for certain outcomes, will have the 
effect of projecting previously separate qubits into a highly 
entangled 
state. This idea has been explored as a route to 
QC using photon qubits 
in a linear optical 
apparatus. Measurement-based gates have 
indeed been shown to be sufficient for universal 
gate-based quantum computation \cite{KLM}. However, in order to achieve 
each logical gate with high probability, one must prepare and then consume 
large auxiliary resources. This necessity is essentially
due to the small probability of success of the 
elementary quantum gates based on auxiliary systems and
measurements \cite{Gates}. 

One way to reduce this overhead is to exploit the idea of 
one-way computing \cite{Old,Long}. 
In this approach 
one would prepare a certain multi-qubit entangled state,
a cluster \cite{Old} or a graph state \cite{Long,Graphs}, 
prior to the computation. This state has the property that the 
computation can then proceed purely by single-qubit 
measurement -- essentially consuming the
graph entanglement as a resource. Recently there has been a successful  
proof-of-principle experiment realising a 4-qubit cluster state 
\cite{ZeilingerNat}. 
A key advantage of the one-way computing strategy is that it 
introduces a degree of separation between the act of creating 
entanglement and the act of
executing the computation. Thus we need not expend the effort needed to 
ensure that each entangling operation
succeeds with high probability -- we can tolerate failures during the 
growth process
simply by rebuilding the affected graph section, provided of course that 
failures are {\em heralded}. Indeed, in this spirit various recent 
schemes \cite{Reznik,MikeCluster,Terry} have
shown how to take gate operations that are
fundamentally non-deterministic, and use them to 
construct an such an entangled resource state 
with certainty.

One particularly attractive possibility
is
to use matter qubits, with the obvious benefits that they are static and  
potentially long lived, together with an optical coupling mechanism 
that creates suitable entanglement. Based on earlier schemes that 
allow for generating entanglement or realizing quantum
gates in matter qubits using flying optical qubits 
\cite{Cabrillo,Plenio,Grangier,Simon,Duan,PlenioNew,Zou,Rempe}, 
two recent publications \cite{Sean,Almut} in particular have
explored precisely this possibility. 
The matter qubits can be completely
separate, for example each within its own cavity apparatus, providing 
that suitable optical channels connect them to a mutual measurement apparatus. 
The simplest scheme is 
that of Barrett and Kok (BK) \cite{Sean}, where one requires only a single beam 
splitter and two detectors in order to couple pairs of qubits. The   
elegant BK approach however suffers from the constraint that, even with 
ideal apparatus, the entangling operation must fail with a probability of
$p=1/2$.
Failures damage the nascent graph state, but because the failure is 
flagged, 
or `heralded', the damaged parts can be removed and the growth can
continue. Nevertheless, the high rate of destructive failures introduces a 
considerable overhead\ \cite{BenjComment}, especially with certain types 
of target graph
topology. The scheme due to Lim, Beige, and Kwek (LBK)\ \cite{Almut} 
introduces the idea of `repeat until success' entanglement, 
meaning that 
while failures  
still occur with probability $p\geq1/2$, these failures are essentially 
passive 
and one can simply try again. Thus one can construct graph    
states with a lower overhead, in terms of number of entangling operations, 
and 
any topology can be directly implemented. However, the cost for 
this advance is that the underlying coupling process is 
more complex: each matter qubit gives rise to a superposition of an 
`early' and a  `late' photon in time-bin encoding, which must subsequently 
enter a beam splitter apparatus simultaneously. This appears to be more
 challenging relative to the simpler BK scheme, 
 so that it is an open 
 question which scheme is the more practical.

Here our goal is to unite the more desirable features of both these 
schemes,
 in particular the simple static optical apparatus of the BK scheme
and the non-destructive `repeat-until-success' aspect of the LBK approach.
 Moreover we introduce a vital feature which neither of these
approaches possess: we demonstrate a graph growth mechanism which does not
 require local unitary operations (e.g., flips) to be performed on the
matter qubits {\em during} the growth process. The growth then becomes 
purely
 a sequence of optical excitations, with a corresponding significant 
increase in speed and considerable reduction in complexity.

We intend that the present paper will form a self contained overview of 
the entire
 paradigm that we are advocating, and to this end we include
compact analysis of the relevant properties of graph states. We make use 
of 
the idea of a minimal graph state (MGS), and make a comparison
with the more limited `cluster states' which results when the geometry of 
physical qubits and their neighbors are fixed by experimental 
constraints. We conclude that there are
dramatic savings, in terms of qubits and entangling operations, when one 
adopts an architecture that can build an MGS directly. 

\section{Graph States and Cluster States}

Graph states \cite{Graphs,Long,SchlingeOld} 
are multi-qubit entangled states, which can be conceived as
having been entangled    
according to certain pattern of two-qubit phase gates. Formally, this 
pattern is specified by the adjacency matrix of an (undirected
simple) graph $G(V,E)$, where $V$ denotes a set of $n$ vertices associated 
with the qubits, and edge set $E$ reflecting the phase gates (see   
Fig.\ \ref{coupling1} (d) for example). The graph state of the empty graph
has the state vector $|\Psi\rangle=
\ket{+}^{\otimes n} = 
((\ket{0}+\ket{1})/\sqrt{2})^{\otimes n}$.
The state vector of a graph state including edges can then be
written as
\begin{eqnarray}
\label{GraphState}
        \ket{G} =
        \prod_{ (a,b)\in E} P^{(a,b)} |+\rangle ^{\otimes n}, 
\end{eqnarray} 
with $P^{(a,b)}$ corresponding to a phase gate
$P^{(a,b)} = (\id + \sigma_z^{(a)} + \sigma_z^{(b)}- \sigma_z^{(a)} 
\otimes 
\sigma_z^{(b)})/2$
between qubits
labeled $a$ and $b$, expressed in terms of Pauli operators. Such 
graph states are stabilizer states \cite{TheStabil}, 
and in turn, every stabilizer
state of $n$ qubits is locally equivalent to a graph state 
\cite{MaartenPhD,SchlingeEquiv}.

A  cluster state  (CS) \cite{Old} is a particular graph state: 
it is one with an underlying cubic lattice of one,
two or three dimensions (see Fig.\ \ref{graph1} (a) for example). 
A cluster state of more than one dimension 
has the remarkable property that it forms a universal
resource for measurement-based one-way computing: having created this 
state, the actual computation is executed simply by making local 
measurements \cite{Long,Old}. It is universal in the sense that 
the procedure amounts to effectively implementing an arbitrary unitary
on the input qubits. 

However, the measurements performed in order to 
implement some chosen algorithm will include
two  classes which it is important to distinguish\ \cite{Long,Old}. The 
first are the Pauli measurements, 
which we can denote as measurements along the $X$, $Y$, or $Z$ axis. Each 
such measurement maps a 
graph state onto another graph state for all outcomes. 
For example 
the $Z$ measurement effectively deletes the measured qubit (node) and its 
associated edges, while $X$ and $Y$ measurements alter the graph according 
to the rules given in Ref.\ \cite{Graphs}. These measurements correspond 
to the Clifford-part of the computation, and the resulting map on the 
level of states can always efficiently be determined on a classical 
computer \cite{Long}. Having performed all the prescribed Pauli 
measurements on a cluster, we are left with a minimal graph state (MGS) 
which is the graph containing the smallest number qubits that is capable 
of realizing our desired algorithm. The remaining measurements are of the 
second class: von Neumann measurements in tilted bases. Such measurements 
take the system out of the graph state prescription and generally cannot 
be efficiently simulated on a classical computer. In a sense one can think 
of the Pauli measurements as simply customizing the (initially universal) 
cluster state into the form that will implement our chosen algorithm, 
while the more general tilted measurements actually execute the algorithm.

Many physical systems that can generate graph states are in fact limited 
to 
cluster state generation, because the physical qubit interactions are 
limited to some kind of nearest-neighbor (or at any rate, local) form. 
This applies to implementations in electron spin lattices and optical 
lattices.
However, we are under no such
constraint since the physical qubits have no defined geometry
\cite{OtherExceptions}. Instead, 
we can directly `grow' an arbitrary graph, and hence
we may prepare the graph state that forms the    
specific resource for a given quantum algorithm. 
We would therefore seek to directly build a MGS, shortcutting the creation 
of the cluster state with its redundant universality.  
This proves to have dramatic advantages in terms of 
the number of entanglement operations and qubits needed. 
In general one finds that a MGS will often exhibit a high
vertex degree, and will be contain significantly 
fewer qubits compared to 
the graph state that is obtained from a cluster
state after measurements along the $Z$ basis, essentially merely removing 
qubits (typically up to an order of magitude). Explicit examples are 
described later.

\section{Description of the Physical Scheme}
In this section we describe the physical requirements and processes involved in implementing our proposal.  We start by describing the  elementary physical systems required.  We then outline the action of the beam splitter network, both on simple product states and more importantly when fusing graph states together. Finally we then show how to do this without any single qubit unitaries, and make some concluding comments. 

\subsection{Physical Components}
\begin{figure}[!h]
  \begin{center}
      \leavevmode
\resizebox{8 cm}{!}{\includegraphics{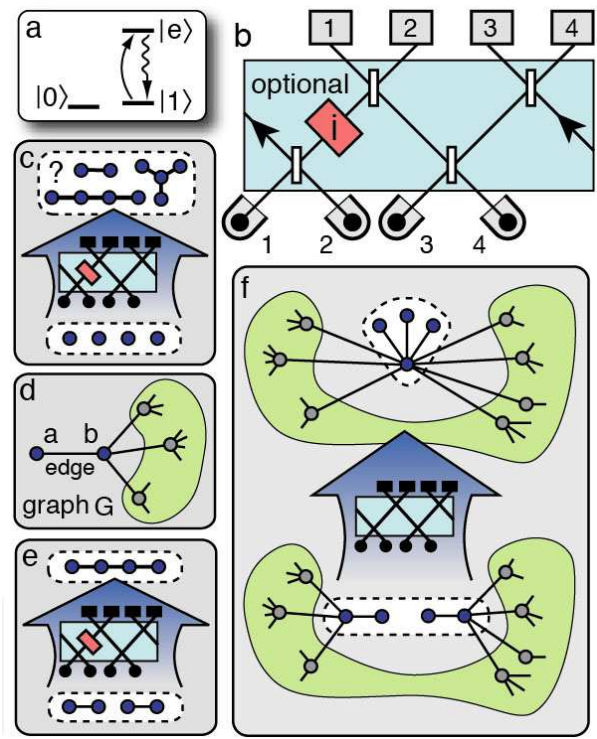}}
\end{center}
\caption{(a) The energy level scheme for a matter qubit. 
(b) Our apparatus: matter qubits $j=1,...,4$ (lower) emit photons to detectors 
$k=1,...,4$ (upper) via 
beamsplitters. We consider two variants of the device: one with a phase 
shifter as marked, one without. (c) The effect of excitation and 
measurement when four {\em product state} qubits are employed -- there are 
various possibly results depending on the number and pattern of detected 
photons. (d) A graph state in which a `leaf' qubit, marked $b$, is 
attached by only one edge. (e) The effect of applying our protocol (with 
shifter) to two EPR pairs. (f) The effect of applying our protocol 
(without shifter) to two arbitrary graphs as shown.}
\label{coupling1}
\end{figure}

In Fig.\ \ref{coupling1} (a) we indicate the basic energy 
level scheme that our matter qubits should incorporate. Obviously, a real 
quantum   
system may have additional levels, but providing these 3-level dynamics are
 incorporated, then such a system is suitable. Candidates
include an actual atom or ion in a trap, of course, but we may also 
consider
 any optically active solid state structure with a discrete
spectrum, such as a quantum dot, or NV-diamond centre 
\cite{sta03}.  The
 ground states, labeled with state vectors $\ket{0}$, $\ket{1}$, are 
the qubit basis states.  The 
third level, labeled $\ket{e}$, 
provides a mechanism for producing a photon from
 the atom, conditional on the atom being in state labeled with 
state vector $\ket{1}$.  That is,
there is an externally driven transition from $\ket{1}\rightarrow\ket{e}$
by using a $\pi$-pulse,
 followed by the optical relaxation $\ket{e}\rightarrow\ket{1}$ which
emits a single photon into the cavity mode, and eventually `leaks' out to
 an external optical network. In a single-mode description, this
is characterized by a coupling strength $\Omega$
of the Jaynes-Cummings coupling between the 
transition $\ket{1}\rightarrow\ket{e}$
and the cavity mode, with decay rate $\Gamma$ of the cavity
mode. The system
is continuously observed via the photon detectors 
placed behind the four beam splitter array (4BS). We note that, as remarked in
Ref.\ \cite{Almut}, if we have a fourth level accessible from
 $\ket{e}$, then we can potentially create a photon directly in the cavity
without significantly populating $\ket{e}$ -- this may be advantageous
 in avoiding dephasing. The state labeled 
$\ket{e}$ is also exploited when we wish
to make a measurement -- continuous illumination by a laser adjusted to
 the transition energy will result in a fluorescence conditional on the  
qubit state. This is a $Z$ measurement; measurement in the other directions
 is accomplished by an appropriate local rotation followed by this  
fluorescence.

The elementary multi-qubit operation in our proposal is based on a  
four-port beam splitter (4BS), which is a composite of four ordinary
 beam splitters, arranged so that every input crosses every other,
 and finally
incident on photon counters. Two types of 4BS will be employed: one
without additional phase shifts, the {\it basic network}, 
and one which includes a 
certain phase-shift corresponding to a factor of 
$e^{i \pi/2}$, the {\it shifter network},
for producing
 certain important cluster states. 
The action of the 4BS is essentially to `erase' information about which
 cavity a photon originated from, so that a given detector cannot
differentiate between matter qubits.  Ideally
the frequencies of the modes are identical, and other sources
of mode matching problems are to a high extent eliminated -- we analyze the 
effects of realistic imperfections in Section VI.

\subsection{Action on Product States}
The analysis to determine the specific projections is straightforward. 
The most simple interesting case we examine is  that of inputting four 
qubits in the {\em product} state corresponding to
$\ket{+}^{\otimes 4}$. The action of the optical 
excitation $\ket{1}\rightarrow\ket{e}$ applied to all qubits, 
followed by the emission
into their local cavities, then results in an equal superposition 
of all basis vectors containing all binary words,
\begin{eqnarray}
	|\phi\rangle = \frac{1}{4}
	\sum_{i,j,k,l=0}^1
	\ket{i,j,k,l}(c_1^{\dagger})^i
	(c_2^{\dagger})^j(c_3^{\dagger})^k(c_4^{\dagger})^l
	|0\rangle .
\end{eqnarray}
where the annihilation operators of the respective
cavity modes are denoted by $c_1,...,c_4$.

As the photons propagate 
through a beam splitter into new modes we employ mappings such as
$(d_j^\dagger,d_k^\dagger)^T=B (c_j^\dagger,c_k^\dagger)^T$, 
$j,k=1,...,4$,
where phases
of the transmitted and reflected mode 
are chosen such that $B$ is given by 
\begin{equation}
	B={1\over\sqrt2}
	\left(
	\begin{array}{cc}
		1 &i \\\ i &1 \\
	\end{array}
	\right).
\end{equation}
Thus the network corresponds to a unitary manipulation of the photon states.
In this way we eventually obtain the
final generation of operators representing photons in the modes 
upon which the detectors act. 
Thus we determine the state that results from a
particular detector reading. In fact a number of states can occur, 
including states that are locally equivalent to linear $4$-qubit 
graphs and
$3$-nodes, as shown in Fig.\ \ref{coupling1} (c). 

\subsection{Fusion of Graph States}
The ability to generate such graphs directly from product states appears 
to  be a promising 
characteristic. However, to properly differentiate the possible outcomes 
one
would require either resolving photon detectors (capable of distinguishing 
a single photon from a pair, etc) or else one would need to resort  
to a lengthy asymptotic variant of the `double heralding' idea in Ref.\ 
\cite{Sean}.
These undesirable features result from the fact that
the systems state, prior to photon measurement, 
was not an eigenstate of the total photon number operator: 
there are elements in the superposition
corresponding to $0,1,...,4$ photons. 
To avoid the problem we must contrive to introduce a 
known number of photons. This is the same
issue faced in Ref.\ \cite{Almut}, where the authors suggest resorting 
time-bin approach and local flips in order to guarantee that each
matter qubit ultimately generates one photon. 

We take a different route, 
based on the idea of {\em fusing} graphs together. We will find that we 
can regard EPR pairs as a kind of raw ingredient from which graphs of 
arbitrary complexity can be grown deterministically.  
Recall that an EPR 
pair with state vector
$|\text{EPR}\rangle = 
(|0,1\rangle - |1,0\rangle)/\sqrt{2}$ is already 
LU-equivalent to the
simplest non-trivial graph state, the one
consisting of two vertices  
connected by an edge. 
We use the term {\em LU-equivalent} to mean,
equivalent up to local unitary operations 
on individual qubits.
Our fusion process exploits existing entanglement 
within the graphs sections: certain 
vertex pairs within a graph state 
can be locally rotated to the subspace
$\text{span} 
\{\ket{0,1}$,$\ket{1,0}\}$ 
-- two such pairs then generate precisely two photons. 

Suppose that a `leaf' node exists, i.e., 
a certain vertex 
(associated with qubit $a$) is attached
to only one other vertex (associated with qubit $b$) of the graph. This is shown in 
Fig.\ \ref{coupling1} (d). Then the
state vector $\ket{\Psi_G} $ of the entire graph state 
is of the following form 
\setcounter{equation}{3}
\begin{equation}
	\ket{\Psi_G} =\left(\ket{0,0}+\ket{1,0}+\ket{0,1}P-\ket{1,1}P\right)\ket{\psi}
\end{equation} 
where the vectors $\ket{a,b}$ correspond to the 
qubits $a$, $b$, and $\ket{\psi}$ 
refers to the external, arbitrarily connected part of the 
graph state, shown inside green bubbles in Fig.\ \ref{coupling1} (d). 
We define 
\begin{equation}
	P=\prod_{i\in N_b\backslash \{a\} }
	\sigma_z^{(i)},
\end{equation} 
so with index $i$ running over the 
neighbors $N_b$ 
of qubit $b$ lying within $\ket{\psi}$. This state vector is in turn equivalent, 
up to a unitary 
rotation 
$(\id-i\sigma_y^{(a)})/\sqrt{2}$ on $a$, to
\begin{eqnarray}\label{LUEQ}
	\ket{\Psi_E} = 
	(\ket{1,0}+\ket{0,1}P)\ket{\psi}
\end{eqnarray} 
Having made this transformation, we know that our qubit pair labeled $a$
and $b$ will emit precisely
 one photon. If we similarly prepare a second pair of qubits,
associated with a different graph (or, a different part of the same 
graph)  then
 the two pairs of qubits can be employed in our 4BS device and
will generate precisely two photons. As indicated in 
Fig.\ \ref{coupling1}, 
we have considered two variants
of the beam splitter network: one with and one without a phase shifter.
A simple analysis determines that in both cases
there are two possible classes of outcome. The two photons may arrive in
a single detector, in which case the effect is simply equivalent to 
applying local
 phase gates to the matter qubits. Alternatively, the two photons
may enter different detectors, in which case the two pieces of graph are
 {\em fused} together in a fashion we specify presently. 
 Identifying the 
various outcomes does not
require counting photon detectors, since we know there are two photons
 in total (but such detectors may be useful in fighting errors when
 taking imperfections into account, see  Section VI). 

The two {\em classes} of outcome are equally probable. In the
 case of the former one can try again without pausing to correct the local
phases, which can be fixed after the eventual successful fusion. The average number of
 attempts required is two. This is then a `repeat-until-success'
scenario equivalent to the one first observed in Ref.\ \cite{Almut}.
  The particular form for our fused graph depends on whether the
phase shifter was employed. If we do employ the shifter,
and supposing that we input two EPR 
pairs, then the resulting state is LU-equivalent to a linear four-qubit graph
as shown in Fig.\ \ref{coupling1} (e). If instead we use our 4BS without
 the phase shifter, i.e. the basic network,
we can couple {\em arbitrary} graph fragments according to
the rule shown in Fig. \ref{coupling1} (f).  For the example of joining two 
EPR states, we show the outcome states conditioned on which detectors 
click, and their probabilities in Table \ref{idealoutcomes}. These 
observations lead us to
 regard EPR pairs as the basic resource for graph growth: EPR pairs can
be generated easily by a single beam splitter using the BK scheme, or
 equivalently we have observed that they can be obtained from our 4BS  
network by choosing to excite just two of the four qubits. The combination
 of the two coupling processes Fig.\ \ref{coupling1} (e) and Fig.\
\ref{coupling1} (f) then allows graphs of arbitrary complexity to be built.
Recall that a graph can be `pruned', i.e., qubit nodes can simply be
removed, by making a $Z$ axis measurement, while other useful transforms
 result from $X$ or $Y$ measurements\ \cite{Graphs}. 
 Indeed, a recent preprint\ \cite{newBrowne}
 makes ingenious use of measurements on leaf structures, reminiscent of 
 those occurring at the fusion point in Fig.\ \ref{coupling1}(f), for qubit loss tolerance 
 in graph states.

 \begin{table}
  \centering 
  \begin{tabular}{ c | c c c c}
$\ket{\Psi(k_1;k_2)}$; Prob    & 1 &2 &3 &4 \\
   \hline
1   &$\ket{a};\frac{1}{8}$ &0; 0 &  
$\ket{c};\frac{1}{16}$&$\ket{d};\frac{1}{16}$\\
 2    & 0;0 &$\ket{a};\frac{1}{8}$ &$\ket{d };\frac{1}{16}$& 
$\ket{c};\frac{1}{16}$ \\
  3   & $\ket{c} ;\frac{1}{16}$&$\ket{d};\frac{1}{16}$& 
$\ket{b};\frac{1}{8}$& 0; 0 \\
  4  & $\ket{d};\frac{1}{16}$ &$\ket{c};\frac{1}{16}$&0; 0 
&$\ket{b};\frac{1}{8}$\\ 
\end{tabular}
  \caption{States vectors resulting from clicks in 
detectors $k_1$ and $k_2$, and 
their probabilites, for the basic beam splitter network acting on the 
state $\ket{\mathrm{EPR}_{1,2}}\ket{\mathrm{EPR}_{3,4}}$, where
  $\ket{a}=(\ket{0,1}+i\ket{1,0})(\ket{0,1}-i\ket{1,0})/2$,  
$\ket{b}=(\ket{0,1}-i\ket{1,0})(\ket{0,1}+i\ket{1,0})/2$, 
$\ket{c}=(\ket{0,1,0,1}-\ket{1,0,1,0})/\sqrt{2}$, 
$\ket{d}=(\ket{0,1,1,0}+\ket{1,0,0,1} )/\sqrt{2}$. State vectors 
$\ket{c}$ and $\ket{d}$ are LU-equivalent to a graph state
vector in which a central vertex radiates 
three `leaf' vertices. }\label{idealoutcomes}
\end{table}

\subsection{Growing a Graph Without Employing Local Gates}
We have seen that we can grow graphs by transforming selected qubits to
 an EPR-type basis prior to fusion, and then applying additional LU
operations to transform the `raw' resultant state back to a graph state. But,
 can we avoid these local transformations? It is evidently necessary to
 employ single qubit rotations at two stages: the very {\em beginning}
 in the entangling procedure,
 where we must take `fresh' qubits and prepare them in $\ket{+}$ in 
 order to
synthesize the EPR pairs which we regard as our basic ingredient, and the
 very {\em end} where we will wish to rotate qubits prior to our
fluorescence measurement, in order to synthesize measurements along some general axis.
 Remarkably, we can in fact omit the numerous local rotations {\em  
during} graph state growth. We find that, within a light constraint on the
 growth process, we can ensure the state remains LU-equivalent to a
graph state at each growth step. 

To see that this is possible consider the
 following argument. 
Suppose that we have some multi-qubit state vector $\ket{\Psi}$ which meets the 
following two conditions:

(i)	\label{induct1}
	The state vector $\ket{\Psi}$ is equivalent up to local unitary
	operations to $\ket{\Psi_G}$ corresponding to 
	a graph 
	$G$ of the form as in Fig.\ \ref{coupling1} (d). 

(ii)
	 Regarding the pair of qubits labeled
	$a$ and $b$, 
	\begin{equation}
	|\braket{0,1}{\Psi}|=|\braket{1,0}{\Psi}|,\,\,	
		|\braket{0,0}{\Psi}|=|\braket{1,1}{\Psi}|=0.
		\end{equation}

From (i) and recalling Eqn.\ 
 (\ref{LUEQ})
we know that $\ket{\Psi}$ is LU-equivalent to a state 
vector $\ket{\Psi_E}=(\ket{1,0}+\ket{0,1}P)\ket{\psi}$ 
since that is itself
LU-equivalent to the graph state vector corresponding to
$G$. The additional constraint (ii) implies that our state vector
can be written as
\begin{equation}
	\ket{\Psi} =(\ket{1,0}U+\ket{0,1}{\tilde U})\ket{\psi}
\end{equation} 
where $U$ is a product of local unitaries acting on $\ket{\psi}$, 
i.e., acting 
in the Hilbert spaces of the qubits {\em other} than those labeled
$a$ and $b$, and ${\tilde U}= \exp(i\phi)UP$ with 
$\phi\in[0,2\pi)$ 
an arbitrary phase. Now let us 
apply the 4BS process of Fig.\ \ref{coupling1} (f) to 
$a$ and $b$, along with an
equivalent pair from some analogous graph state 
(or, another part of the same
 graph state). 
 On failure (with probability $1/2$) 
 the process simply introduces some known local phases, which do not alter 
our prescription. On eventual success we 
 generate a state vector
 \begin{equation}
	\ket{\Psi_{\rm tot}} = 
	(\ket{X}U_{\rm tot}+\ket{{\bar X}}{\tilde U}_{\rm tot})\ket{\psi_{\rm tot}}.
\end{equation} 
Here $X$ is a binary word with two zeros, two ones, and ${\bar X}$ is its complement. 
The vector 
$\ket{\Psi_{\rm tot}}$ refers to 
the entire state vector of the fused system, 
and $\ket{\psi_{\rm tot}}$ is the state vector for all 
qubits except the four that coupled via the 4BS. 
Similarly, $U_{\rm tot}$ is some product of local unitaries 
on $\ket{\psi_{\rm tot}}$, and 
${\tilde U}_{\rm tot}= \exp(i\theta)P_{\rm tot}U_{\rm tot}$, 
where 
\begin{equation}
	P_{\rm tot}=\prod_i \sigma_z^{(i)}
\end{equation}	
with the index running over neighbors of either of the original 
vertices (but excluding mutual neighbors). 
The phase $\theta\in[0,2\pi)$ is determined by  
$\phi$ and its counterpart in the second pair, together with phases introduced in any 
failures preceding the successful fusion. 
This state vector $\ket{\Psi_{\rm tot}}$ is indeed 
LU-equivalent to the desired fused graph state of 
Fig.\ \ref{coupling1} (f). 
Moreover, because $X$ has two zeros and two 
ones, if we nominate one of those four qubits to be
a new `vertex' qubit $b$, two of the three 
remaining `leaf' qubits are available to be
 labeled as $a$ to satisfy (i) and (ii). Thus we can go on to perform further fusions 
using $\ket{\Psi_{\rm tot}}$. 
 To conclude the argument we need only observe that conditions (i) and (ii) are  
met by simple EPR pairs, and
by the state 
 resulting from fusing two EPR pairs via the process depicted in Fig.\ \ref{coupling1} (e), i.e.
 the state that is LU-equivalent to a linear four qubit graph state. 
Thus these simple states can act as the initial building blocks as we 
construct a complex graph.

Then provided we are prepared to  measure out
 one in every three of the `leaf' 
nodes which occur at each fusion point, we
can grow our entire graph from EPR pairs without the use of local unitary operations
during the growth process. This
 constraint is extremely light: we would rarely wish to use all  
three leaves, and in any case the number of leaves can be increased by two
 simply by fusing an EPR pair, which adds one to the number of leaves
that are eligible in the sense of property (ii). Of course, the LU 
operations
 needed to map the final state to the desired ultimate graph state
can subsumed into the rotation which in any case precedes measurement. The
 growth process is therefore entirely one of optical excitation and 
detector 
monitoring. One can anticipate than in many systems, the cost in efficiency
 arising from following the constraint would be vastly outweighed by
the increase in growth speed.

\subsection{Further Remarks on Graph Growth}

There is one additional comment to make regarding the speed of our 
protocol: in the
scheme of Ref.\ \cite{Sean} it is always necessary to wait a period 
after the initial measurement to ensure there are no further
photons in the apparatus. The fidelity of the entangled states is only 
high if this wait period is long
compared to the typical time for a photon to be detected. 
This additional
waiting time necessary in Ref.\ \cite{Sean} is not necessarily long, 
given that photon emission from these sources is approximately 
exponential, governed by the time scale $1/\Gamma_{\text{Slow}}$. 
By contrast, 
because we contrive to have precisely 
two photons in the apparatus, once we see two detection events (either in 
different detectors, or, given resolving detectors as discussed later, 
within one detector) we have no need for such a wait. 
One should hence expect a factor of about $5$
in difference concerning the
speed of this step in the respective schemes.

In a mature form of the architecture described here, one would envisage
 coupling the many qubits by a form of $N$-port all-optical router, 
as used in {\it integrated optics},
so that our qubits can remain static 
and we can choose which of them will
 couple by suitably setting the router and optically exciting only that
subset. Devices relevant to this technology have already been developed
 for classical optical communications \cite{BellLambda}. This would permit
direct growth of graphs with an arbitrary topology, and in particular the
 ability to directly entangle 
arbitrary qubits gives a 
non-local architecture \cite{NL1,NL2,NL3}.
Such non-local architectures 
may prove to have an 
advantage in quantum fault tolerance and error
correction \cite{Svore}.

\section{Characteristics of the derived graph states}

As discussed above and illustrated in Fig.\ \ref{coupling1}, our protocol 
can generate graphs of arbitrary topology, including nodes of high degree. 
We argued earlier in Section II that, given such an architecture, 
one should 
aim to create minimal graph states (MGS). The advantage in terms of 
resource consumption when preparing appropriate
 MGS compared to standard cluster states (CS) can be quite significant. 

\begin{figure}[!h]
  \begin{center}
    \leavevmode 

\includegraphics[width=7cm]{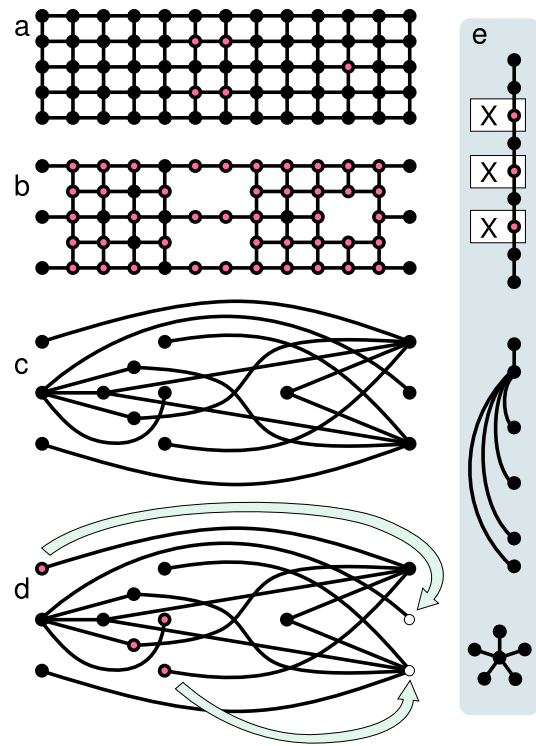}

\end{center} 
\caption{Progressive measurement on qubits (marked
red). (a) A fully connected cluster state. After $Z$ measurements (b) and 
$X$ and $Y$
 measurements we obtain a minimal graph state (c) for the
algorithm, here a Toffoli gate. In practice one would further reduce the 
number of
 qubits required by just in time graph creation -- any
qubit whose set of edges are complete can be measured and `recycled' to 
the `front'.
 Side panel (e) indicates the overhead required in the BK
scheme in making high degree vertices: one measures out a portion of the 
qubits (in the X basis here) in order to alter the graph topology.}
\label{graph1} \end{figure}
  
 \subsection{Minimal Graph States}
  
 In the best known   
scheme for an $n$-qubit quantum Fourier transformation
\cite{Long}, the number of
 required qubits is 
	$C_{\text{Fourier}}(n)= 8 n^2 + O(n) $
for a cubic
cluster, followed by first $X$, $Y$, and $Z$ measurements, then tilted
 measurements.
 The MGS embodies 
	$G_{\text{Fourier}}(n) =
	(3/2) n^2 +O(n)$
 qubits. In turn, for the quantum adder, 
\begin{eqnarray}
	C_{\text{Adder}}(n) = 312 n+O(1),\,\,
	G_{\text{Adder}}(n) = 16n +O(1)
\end{eqnarray}
\cite{Long}. Hence, one may gain more than an order of magnitude in 
resource consumption.
For the $3$-qubit Toffoli gate that we use here as our illustration, 
we have $G_{\text{Toffoli}} = 13$ versus $C_{\text{Toffoli}} = 65$, so a 
factor of $5$ difference in the number of qubits.

\subsection{Generation of Edges}

Concerning the actual preparation of the graph states,
we emphasise two points: firstly, when introducing edges with a physical
interaction, one should always prepare the LU-equivalent
graph state corresponding to the graph with the minimal number of 
edges. Or, more specifically, the graph state with the minimal number
of edges that is equivalent
up to local Clifford unitaries \cite{Remark}, 
which merely amount to a local Clifford 
basis change. This has
also been emphasised in Ref.\ \cite{Perdrix}.
Fortunately, an efficient algorithm is known to 
check full local Clifford equivalence \cite{MaartenEquivalent}.
Any graphs that correspond to local Clifford equivalent graph states
can be related to each other with a successive application
of local complementations \cite{Graphs,MaartenEquivalent}.
Also, it is known how many different graph states are contained 
in an equivalence class with respect to 
local Clifford unitaries \cite{MaartenPhD}.

Secondly, the present scheme 
seems particularly suitable to prepare graph states
of graphs involving vertices with a high vertex degree in
a single step. In a CS after
measurement of the unused qubits along the $Z$ direction, it suffices to 
have vertices with a maximal vertex degree of $3$. This is 
obviously the lowest possible for a graph less trivial than a 
linear cluster state. 
From the example of Fig.\ \ref{graph1} one would suspect 
that a typical MGS may {\em need} higher degree nodes, and this is an 
important question in considering how they can be efficiently 
constructed. To explore this point we consider the 
`maximal vertex degree', by which we   
mean, the highest degree of any vertex in the graph. 
The vertex degree in a MGS can in principle take any value. 
The maximal vertex degree is notably not invariant under 
local Clifford unitaries \cite{Remark}.
To render the notion of maximal vertex degree meaningful, we have to
take its minimum value when minimized over all local
Clifford unitaries. For a GHZ state of $n$ qubits, it can 
easily be shown that it has a smallest 
maximal vertex degree of $n-1$. 
So we immediately see that it is meaningful to talk 
about `highly connected graph states'. 
For the resource state for the $[7,1,3]$-CSS code as considered in Ref.\
\cite{Graphs}
 we find that in the whole orbit under local Clifford unitaries the 
smallest
maximal vertex degree is $6$, whereas the largest is $34$. In the
 three-qubit quantum Fourier transform the smallest maximal vertex degree 
is
$4$. Thus we see that high degree
vertices are indeed generally unavoidable in a MGS: any
scheme that claims to be able to directly and efficiently construct
 such a state must be able to create graph states with 
vectices of high vertex degree. 
The BK scheme, for example, 
appears somewhat limited by the increased difficulty of making the high
	degree vertices associated with graph states; the high rate
	of destructive failures leads one to take an indirect approach as depicted in 
	Fig.\ \ref{graph1}(e), with an associated cost in resources. The
scheme presented here is among the few that generate high degree nodes directly
\cite{OtherExceptions}, as one can quickly 
see by considering the fusion rule
depicted in Fig.\ \ref{coupling1}(f).

\subsection{Cluster States}
       
Of course, if for some reason we wished to 
generate a conventional CS 
rather than the
MGS specific to some given algorthim, then we can do so efficiently. As an 
exercise, let us conclude this section
with a comment on the  required
number of steps in the preparation of a CS
with an underlying two-dimensional cubic lattice.
We will count resources in terms of the number of
applications of the 
shifter $N_{\text{Shifter}}$, with two EPR pairs fed in in each 
instance, and of  the basic network $N_{\text{Basic}}$. The basic
building block can be taken to be a cross shape of length
$4$, requiring four EPR pairs, and the application of 
two shifter and one basic networks.  One row of width $n$
can be build using $2n$ invokations of the shifter
network and $2n$ uses of the basic network. 
A two-dimensional cluster state
on a $n\times n$ cubic lattice hence requires
\begin{equation}
N_{\text{Shifter}}(n) =2n^2,\,\,
N_{\text{Basic}}(n)=3n^2-n.
\end{equation}

\section{Just in Time Graph Creation}

In a lattice system, one may be well advised to prepare the 
multi-particle cluster state in one step, exploiting a natural 
nearest-neighbor
interaction. However, in a scheme as considered here, there is 
no motivation to prepare graph edges far in advance of the eventual 
measurement
operations that will consume them. One should therefore avoid doing so 
since this gives rise to unnecessary errors due to the graduate
degradation in phase integrity from decoherence. Instead, one can 
introduce new edges and vertices for our MGS 
shortly before it is needed, in
a manner analogous to the block-by-block process of Ref.\ 
\cite{Terry} but 
at a finer scale. By analogy to the term used in classical computing, 
this may be referred to as {\em just in time} graph state generation
\cite{Prehist} (see also Refs.\ \cite{Long,Elham}).
As noted earlier in Section III, although we may require local unitaries 
to create the EPR pairs which constitute our `raw ingredient', the
remaining steps involved in generating new graph structure can take place 
without such manipulations.

One can easily confirm that this is possible, 
even though the measurements
 on earlier parts of the graph are tilted and therefore have taken
the system to a non-graph state. Consider a graph state with graph 
$G=(V,E)$ corresponding to the  whole  computation: let us consider the 
state   vector after 
$k$
measurements on vertices $ a_1,...,a_k $, forming a vertex set $V_k\subset 
V$. The resulting state vector after measurements 
in direction $r_k$  -- depending 
on the measurement outcomes $s_1,...,s_{k}\in\{-1,1\}$,  -- in this 
temporal order 
is given by 
        $P_k |G\rangle = \prod_{j=1}^k ({\id + (-1)^{s_j} r_j
        (s_1,s_2,...,s_{j-1})\sigma^{(a_j)} )/2)}|G\rangle$,
where $\sigma^{(a_j)}$ is the vector of Pauli matrices at vertex
labeled $a_j$. Note that 
the appropriate measurement basis $r_j
        (s_1,s_2,...,s_{j-1})$ at step $j$ depends on the
earlier measurement outcomes. 
Yet, at this point we could have just prepared
\begin{eqnarray}
        \ket{G_k} =
        \prod_{ (a,b)\in E_k} P^{(a,b)} |+\rangle ^{\otimes n}
\end{eqnarray} 
before performing the above measurements, 
where
$       E_k=
        \left\{
        (a,b)\in E: a\in V_k \text{ or } b\in V_k
        \right\}$.
Thus the only constraint on this just in time approach, 
is that one should ensure that all edges in $E_k$
are appropriately entangled in step $k$, see Fig.\ \ref{graph1} (d).

\section{Error Analysis}

A physical implementation of this scheme would be subject to a number of
possible errors.  Our protocol relies on the subsystems being identical,
so that their outputs are indistinguishable.  Thus, mismatching parameters
will lead to a reduction in performance.  Other errors include dephasing
of the matter qubits, imperfect optical excitation, 
phase noise (or drift) in the optical apparatus, and photon loss.  section
the majority of our analysis will focus on errors due to mismatched subsystems; we 
will comment on the other error sources at the end.

Since the results described here involve the detection of two photons
arriving from a source, the qualitative effect of errors will be similar
to the results presented in Ref.\
\cite{sta03}, and we analyse the system using
similar methods.   For the purpose of this analysis, we assume that each
atom is a three level system, with degenerate ground states,
labeled $\ket{0}$ and
$\ket{1}$, and a level $\ket{e}$ that is optically coupled to $\ket{1}$,
with an energy ${\hbar\omega_e}$.  The cavity is taken to have a frequency
$\omega_c=\omega_e+\Delta$, where $\Delta$ is nominally zero.  The
transition $\ket{e}\leftrightarrow\ket{1}$ couples to the cavity mode with
a strength $\Omega$, and the cavity mode decays with a rate $\Gamma$.
Thus, we consider here imperfections in $\Delta_j$, $\Omega_j$ and
$\Gamma_j$ for each atom-cavity subsystem, $j=1,...,4$. A comparable analysis was performed
in Ref.\ \cite{Sean}, so that we can compare that two-qubit scheme with the
present four-qubit protocol. Remarkable, we find that the sensitivity to defects in the apparatus is
essentially the same.

\subsection{Continuous Measurement Analysis}

In the following we describe the
dynamics of a three-level atom in a leaky cavity
in the Schr{\"o}dinger picture, continuously
monitored by a photodetector. 
Its stochastic dynamics under continuous measurement can be
described using a quantum-jump approach,
leading to a piecewise deterministic classical stochastic
process in the set of all pure states \cite{gar00,Knight,Holevo}.
The continuous time evolution is governed by an
effective Hamiltonian, interrupted by discontinuous `jumps'
reflecting photon detection.
This continuous part is described by the
Schr\"odinger equation
$\partial_t\ket{\tilde\psi}=-i \heff \ket{\tilde\psi}$ for the
unnormalised state vector $\ket{\tilde\psi}$,
with the non-Hermitian, effective Hamiltonian
\begin{eqnarray}
\heff=\omega_e \ket{e}\bra{e}+(\omega_c-i\Gamma/2) c^\dagger c+\Omega
(c\ket{e}\bra{g}+c^\dagger\ket{g}\bra{e}),\nonumber\\
\end{eqnarray}
where $c$ is an annihilation operator for the cavity mode.
The decreasing vector
norm of $\ket{\tilde\psi(t)}$ due to the non-unitary
evolution, $\ueff(t)=e^{-i \heff t}$, leads to the cumulative density
function for the time at which the photodetector registers a photo-count,
\begin{equation}
P(t)=1-||\,\ket{\tilde\psi(t)}||^2.
\end{equation}
This in turn governs the waiting
time distribution in the stochastic process.
Correspondingly, a detector click
corresponds to a `jump' in the state of the system according to
$\ket{\psi}\longmapsto
\gamma c\ket{\psi}$, where $\gamma=\Gamma^{1/2}$ {\cite{gar00}}.

\begin{figure}
  \begin{center}
\includegraphics[width=4cm]{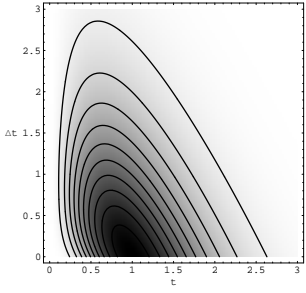}
\end{center}
\caption{Probability density function, $p(t;\Delta t)$, for measuring the first photon at time $t$ followed by the second after a delay of $\Delta t$.  We have used
$\Omega=1$, $\Gamma=4$ and $\Delta=0$ for all subsystems.}
\label{prob}
\end{figure}

For the system of four atoms in cavities, with four detectors following a
beam splitter network, the only change to this prescription is that
$\ueff(t)=\prod_j\ueff_j(t)$ and a click in detector $k$ effects a `jump'
in the state according to $\ket{\Psi}\longmapsto d_k\ket{\Psi}$, where
$d_k$ is related to the 
cavity mode operators according to, $d_k=\sum_j
\beta_{k , j}\gamma_j c_j$, where $\beta= (\beta_{k,j})$ is 
the unitary induced by the beam splitter
network.  Note that we use $j=1,...,4$ to
label subsystems and $k=1,...,4$ to label detectors.

We examine the effect of errors on the basic 4BS network which creates
4-GHZ states from two EPR pairs, so that the initial state
vector is given by
$\ket{\Psi(0)}=\ket{\mathrm{EPR_{1,2}}}\ket{\mathrm{EPR_{3,4}}}$.
We treat
the initial excitation of the protocol, $\ket{1}\rightarrow\ket{e}$, as
instantaneous and ideal, and examine the effect of mismatched system
parameters $\Gamma_j$, $\Delta_j$ and $\Omega_j$ on the subsequent emission and detection process.  The
attraction of our protocol is that the initial state is in the
two-excitation subspace, so we expect to register exactly two detector counts.
The state vector of the system at the end of the protocol is conditional upon
which detectors clicked, $k_1$ and $k_2$, and at what times, $t_1$ and
$t_2=t_1+\Delta t$,
\begin{equation}
\ket{\tilde\Psi(t_1,k_1;t_2, k_2)}=d_{k_2} \ueff(\Delta
t)d_{k_1}\ueff(t_1) \ket{\Psi(0)}.
\end{equation}
This particular outcome occurs with a probability density function given
by $p(t_1,k_1;t_2, k_2)=|| \,\ket{\tilde\Psi(t_1,k_1;t_2, k_2)}||^2$.  Integrating $p$ over $t$ and $\Delta t$ yields the probabilities in Table \ref{idealoutcomes}, as required.  Note that for the \emph{ideal} case, the distribution of $p$ does not actually depend on $k_1$ and $k_2$; the only dependence is an overall multiplicative factor such that Table \ref{idealoutcomes} is satisfied.  An example is shown in 
Fig.\ \ref{prob}.

For a given combination of detectors and times, we calculate the fidelity
of the resulting state with respect to the ideal outcomes, shown in Table
\ref{idealoutcomes}, which depend only on the detectors, and not the
times: $f(t_1,k_1;t_2, k_2)=|\langle{\Psi(k_1,k_2)}\ket{\Psi(t_1,k_1;t_2,
k_2)}|^2$ (note that we have renormalised the state, denoted by the lack
of a tilde). In order to give a fair estimate of the expected fidelity,
we compute the time-averaged fidelity for outcomes where different
detectors click,
\begin{eqnarray}
F&=&2\sum_{k_1\neq k_2}\int_0^\infty \hspace{-3mm}dt_1 \int_{t_1}^\infty
\hspace{-3mm}dt_2\, p(t_1,k_1;t_2, k_2) f(t_1,k_1;t_2, k_2),\nonumber\\
&=&2\sum_{k_1\neq k_2}\int_0^\infty \hspace{-3mm}dt_1 \int_{t_1}^\infty
\hspace{-3mm}dt_2\, \tilde f(t_1,k_1;t_2, k_2),
\end{eqnarray}
where $\tilde f(t_1,k_1;t_2, k_2)=|\langle{\Psi(k_1,k_2)}\ket{\tilde
\Psi(t_1,k_1;t_2, k_2)}|^2$, and the factor of 2 accounts for the fact that in
the ideal case the probability of fusing the states is $1/2$.  $\tilde f$ consists of a sum of exponentially decaying terms, so we compute
$F$  analytically, though the expression is rather lengthy.

\begin{figure}
  \begin{center}
\includegraphics[height=4.2cm]{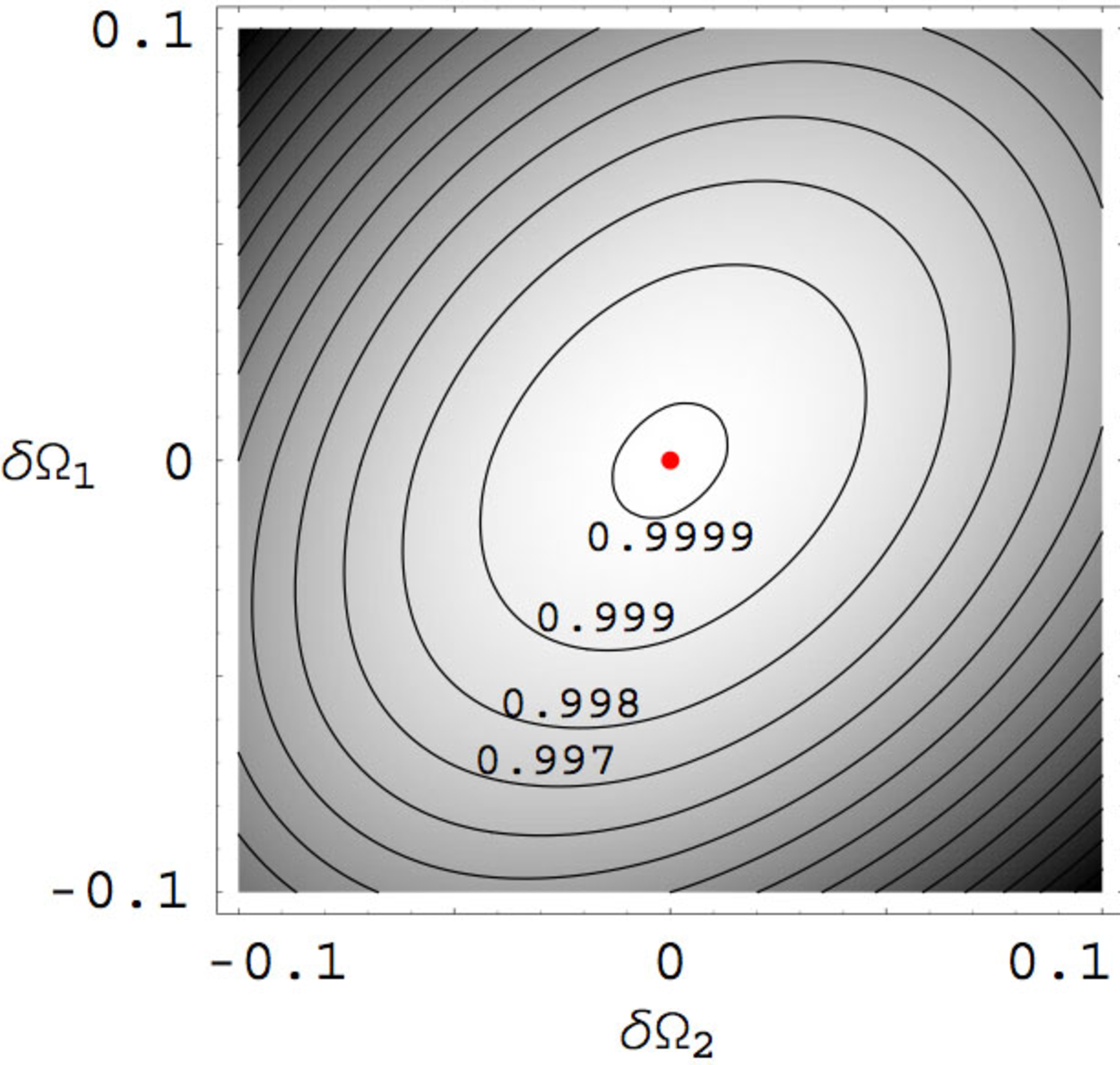}
\includegraphics[height=4.2cm]{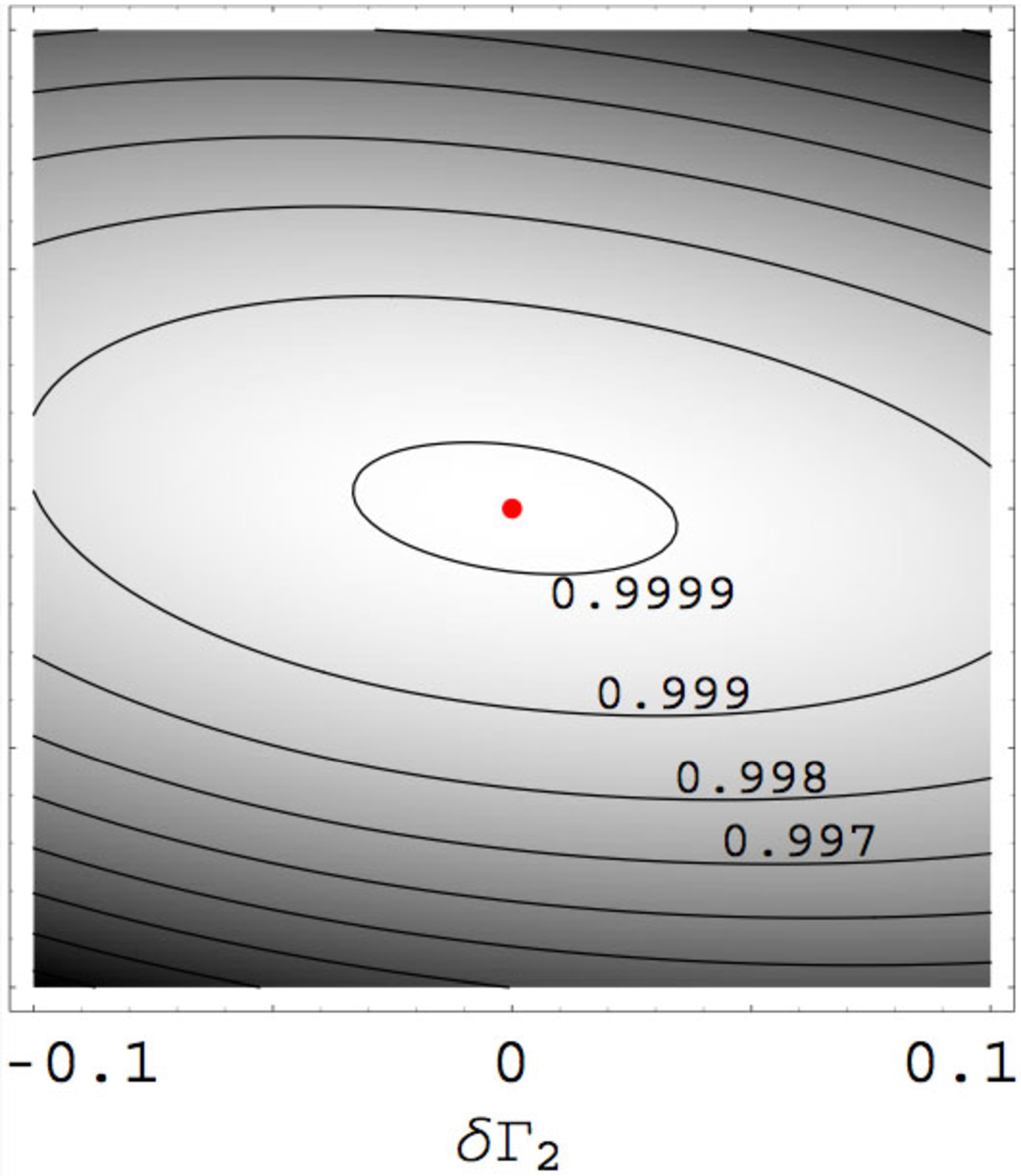}
\end{center}
\caption{Contours of fidelity, $F$, as a function of $\Omega_{1}=1+\delta\Omega_{1}$ (vertical axes) and
$\Omega_{2}=1+\delta\Omega_{2}$ (left), and  $\Gamma_{2}=4+\delta\Gamma_{2}$  (right).  All
other parameters are fixed at $\Delta_j=0$, $\Omega_j=1$ and
$\Gamma_j=4$.  The innermost contour is at $F=1-10^{-4}$ whilst all others
are at $10^{-3}$ intervals.}
\label{fidelity}
\end{figure}

When  $\Omega_j=\Omega$ and $\Gamma_j=\Gamma$, we find $F=1$, so that the
protocol works perfectly.  Otherwise, the process works with reduced
fidelity, as shown in Fig.\ \ref{fidelity}.  Since we are considering
three error parameters per subsystem, we cannot show the dependence of $F$ on all of them
graphically, however for small perturbations to the parameters, we can
straightforwardly compute the dependence to quadratic order.  In what
follows, we work in units for which $\Omega=1$.  We find that
$F\approx1-\sum_j{\epsilon}_j^{\,\,T}
M_s {\epsilon}_j-\sum_{j_1\neq j_2}{\epsilon}_{j_1}^{\,\,T}
M_\times {\epsilon}_{j_2}$, where 
\begin{equation}
{\epsilon}_j=(\delta
\Delta_{j},\delta \Gamma_{j},\delta \Omega_{j})
\end{equation}
is the vector of
parametric errors in subsystem $j$ and $M_{s,\times}$ are the coefficient
matrices.   For a critically damped atom-cavity system, 
$\Gamma=4\Omega$,
the coefficient matrices are
\begin{eqnarray}
M_s=\left[\begin{array}{rrr}\frac{5}{128} & 0 & 0 \\0 & \frac{3}{32} &
-\frac{3}{16}  \\0 & -\frac{3}{16}  & \frac{9}{16}
\end{array}\right],\,
M_\times=\left[\begin{array}{rrr}-\frac{3}{128}  & 0 & 0 \\0 &
-\frac{1}{32} & \frac{1}{16} \\0 & \frac{1}{16} &
-\frac{3}{16}\end{array}\right].\nonumber\\
\end{eqnarray}

\subsection{Photon Loss}

Given an ideal
apparatus, without any photon loss or detector failure, our fusion process would merely require four simple non-photon-number-resolving detectors. However, in practice any
near-future experiment will certainly suffer significant photon loss. This appears potentially very damaging to our scheme (and to
that of Ref.\ \cite{Almut}, but not to that of Ref.\ \cite{Sean}),
because we may misinterpret a photon loss as two photons
entering a single detector.
In order to counter this issue, we would require a limited degree of photon resolution at the detectors -
specifically, we must differentiate the three cases: $0$, $1$ and {\em more than one} photons. This suffices to detect a photon loss event -- we would then reset the associated matter
qubits and rebuild the graph section (analogously to Ref.\ \cite{Sean}). 
Importantly, graph state fidelity will not be affected by undercounts, reflecting a non-unit detection
efficiency, which is the dominant problem in real world photon
detector technologies. Undercounting is equivalent to a lossy channel followed by a
perfect photon counter, and therefore detector inefficiency is simply subsumed into the total photon loss rate.  
Detector over-counting, i.e. dark counts, are potentially harmful in the present scheme and those 
of Ref.\ \cite{Sean,Almut}. Fortunately, since we know that correct operation of the scheme 
generates precisely two photons, we will successfully identify any photon loss event unless the loss
occurs {\em at the same time} that a detector is
independently subjected to a dark count, a process that
is expected to happen with a very small probability. 

One could construct an adequate detector simply from two non-photon-resolving detectors, together with a fast switch. This would exploit that fact that when two photons are incident on
one detector, they are typically separated by a time interval of order
$1/\Gamma$, see Fig.\ \ref{prob}; this time
can be made long enough to trigger a
pockels cell to redirect a possible
second photon into a second detector,
obviating the need for a true number-resolving detector. On occasions when the two photons occur too close together for the second to be redirected, we simply undercount and assume photon loss has occurred.

To summarize, the present scheme is potentially more susceptible to photon loss than the `double heralding' scheme of Ref.\ \cite{Sean}. However, the issue can be dealt with using detector technologies that remain relatively simple -- we do not require high fidelity photon number resolving detectors in order to generate high fidelity graph states.

\subsection{Other Errors}

This shows that the protocol is most sensitive to errors in the
atom-cavity coupling rate, and less sensitive to detuning or the cavity
leakage rate (the same hierarchy as observed in Ref.\ \cite{Sean}) .  The method used here can be adapted to include dephasing
errors, as it was in Ref.\
\cite{sta03}, however it is rather more cumbersome,
so for brevity we do not analyse it in detail here.
In Ref.\ \cite{sta03}, it
was found that dephasing was minimised when $\Gamma\approx\Omega$, since
such a critically damped system  has the shortest lifetime of excitations
in the system.  It was also noted that dephasing was negligible when
$\Omega$ and $\Gamma$ are much larger than the dephasing rate.  We expect
these statements to hold true in this system as well, since the underlying
physical processes are the same.

The issue of {\em interferometric instability} is relevant to any scheme in which terms in the matter qubit superposition become 
coupled to the presence/absence of a photon in a given channel. Any phase noise suffered by a photon in transit through the apparatus ultimately can be mapped onto the matter qubits. 
Fortunately, there has been enormous progress recently
in the development of experimental techniques for phase
locking, which should prove to be beneficial for a
scheme of the proposed type\ \cite{Gilchrist}.
Imperfect optical excitation can be dealt with by noting that this simply
reduces the  initial state fidelity, which thus reduces the protocol
fidelity by an equal amount.

\section{Summary}

We have described a scheme that unifies some of the desirable 
features of previous
 work on matter qubits and graph states. It is able to achieve
{\em deterministic} growth while using simple static 
linear optics and a `one shot' excitation. Moreover, 
the presented scheme obviates the need for
continual local operations on qubits during graph growth, which 
implies a dramatic speedup in many systems. 
The scheme proves to have
properties that make it ideal for creating the most resource
efficient form of graph state, the minimal  graph state. These
minimal graph states, which form the essential resource 
for a given quantum computation, without its classically efficiently
tracktable Clifford-part, typically 
correspond to graphs with a high maximal vertex degree. 
For the preparation of such graph states this scheme
is particularly suitable.  We observe that the use of minimal graph states 
is completely   
compatible with the idea of `just in time' entanglement 
generation. 
Our protocol is relatively robust to mismatch in the subsystems, and an 
accuracy of greater than $1\%$ in the parameters will provide a fidelity 
of around $0.9999$ in the final state. 
We hope that the  scheme presented in this 
work can contribute to bringing 
full-scale graph state quantum computation
closer to practical realization.

\section{Acknowledgements}
We would like to thank 
H.J.\ Briegel,
D.E.\ Browne, 
W.\ Munro, and
E.\ Solano 
for fruitful discussions, and S.\ Barrett,
P.\ Kok, and M.B.\ Plenio for
helpful comments on the manuscript.
This work has been supported by the 
EPSRC (QIP-IRC), the
EU (IST-2002-38877), 
the DFG (Schwerpunktprogramm QIV), the European
Research Councils (EURYI) and the Royal Society.


\begin{thebibliography}{99}

\bibitem{Nielsen}
	M.A.\ Nielsen and Chuang, {\it Quantum computation}
	(Cambridge University Press, Cambridge, 2000).

\bibitem{Blatt}
	F.\ Schmidt-Kahler, H.\ H{\"a}ffner, M.\ Riebe, 
	S.\ Gulde, G.P.T.\ Lancaster, T.\ Deuschle, 
	C.\ Becher, C.F.\ Roos, J.\ Eschner, and 
	R.\ Blatt, Nature {\bf 422}, 408 (2003).

\bibitem{Wineland}
	D.\ Leibfried, B.\ DeMarco, V.\ Meyer, D.\ Lucas, M.\ Barrett,
	J.\ Britton, W.M.\ Itano, B.\ Jelenkovic, C.\ Langer, 
	T.\ Rosenband, 	
	and D.J.\ Wineland, Nature {\bf 422}, 412 (2003).

\bibitem{NVDiamond}
	F.\ Jelezko, T.\ Gaebel, I.\ Popa, A.\ Gruber, 	
	and J. Wrachtrup,  
	Phys.\ Rev.\ Lett.\ {\bf 92}, 076401 (2004).

\bibitem{NVDiamond2}
	M.\ Riebe, H.\ H{\"a}ffner,
	C.F.\ Roos, W.\ H{\"a}nsel,
	J.\ Benhelm,
	G.P.T.\ Lancaster,
	T.W.\ K{\"o}rber,		
	C.\ Becher, F.\ Schmidt-Kaler, D.F.V.\ James,
	and R.\ Blatt, Nature {\bf 429}, 734 (2004).

\bibitem{Forster}
	A.\ Nazir, B.\ Lovett, S.\ Barrett, J.H.\ Reina, 
	and A.\ Briggs, Phys.\ Rev.\ B {\bf 71}, 045334 (2005).

\bibitem{KLM}
        E.\ Knill, R.\ Laflamme, and G.\ Milburn, Nature {\bf 409},
	46 (2001).

\bibitem{Gates}
        J.\ Eisert, Phys.\ Rev.\ Lett.\ {\bf 95},
040502 (2005); S.\ Scheel and
        N.\ Luetkenhaus, New J.\ Phys.\ {\bf 6}, 51 (2004);
        E.\ Knill, Phys.\ Rev.\ A {\bf 68}, 064303 (2003).

\bibitem{Old}
        R.\ Raussendorf and H.J.\ Briegel,
        Phys.\ Rev.\ Lett.\ {\bf 86}, 5188 (2001).

\bibitem{Long}
        R.\ Raussendorf, D.E.\ Browne, and H.J.\ Briegel,
        Phys. Rev. A {\bf 68}, 022312 (2003).

\bibitem{Graphs}
        M.\ Hein, J.\ Eisert, and H.J.\ Briegel,
        Phys. Rev. A {\bf 69}, 062311 (2004).
        
\bibitem{ZeilingerNat}
	P.\ Walther, K.J.\ Resch, T.\ Rudolph, E.\ Schenck, 
	H.\ Weinfurter, V.\ Vedral, M.\ Aspelmeyer, and A.\ Zeilinger,
	Nature {\bf 434}, 169 (2005).
	
\bibitem{Reznik}
	N.\ Yoran and B.\ Reznik,
	Phys.\ Rev.\ Lett.\ {\bf 91}, 037903 (2003).

\bibitem{MikeCluster}
	M.A.\ Nielsen, Phys.\ Rev.\ Lett.\ 
	{\bf 93}, 040503 (2004).

\bibitem{Terry}
        D.E.\ Browne and T.\ Rudolph, Phys.\ Rev.\
Lett.\ {\bf 95}, 010501 (2005).

\bibitem{Cabrillo}
	C.\ Cabrillo, J.I.\ Cirac, P.\ Garchia-Ferndandez, 
	and P.\ Zoller,
	Phys.\ Rev.\ A {\bf 59}, 1025 (1999).	

\bibitem{Plenio}	
	S.\ Bose, P.L.\ Knight, M.B.\ Plenio,
	and V.\ Vedral, Phys.\ Rev.\ Lett.\
	{\bf 83}, 5158 (1999).

\bibitem{Grangier}	
	I.E.\ Protsenko, G.\ Reymond, N.\ Schlosser, and P.\ Grangier,
	Phys.\ Rev.\ A {\bf 66}, 062306 (2002).	

\bibitem{Simon}
	C.\ Simon and W.T.M.\ Irvine, 
	Phys.\ Rev.\ Lett.\ 		{\bf 91},
	110405 (2003).

\bibitem{Duan}	
	L.M.\ Duan and H.J.\ Kimble, Phys.\ Rev.\ Lett.\ {\bf 90},
	253601 (2003). 	

\bibitem{PlenioNew}
	D.E.\ Browne, M.B.\ Plenio, and S.\ Huelga, 
	Phys.\ Rev.\ Lett.\ {\bf 91}, 067901 (2003).

\bibitem{Zou}
	X.B.\ Zou and W.\ Mathis,	
	Phys.\ Rev.\ A {\bf 71}, 042334 (2005).

\bibitem{Rempe}
T.\ Legero, T.\ Wilk, M.\ Hennrich, G.\ Rempe, and
A.\ Kuhn, Phys.\ Rev.\ Lett.\ {\bf 93}, 070503 (2004).

\bibitem{Sean}
        S.D.\ Barrett and P.\ Kok, Phys.\ Rev.\ A
{\bf  71}, 060310(R) (2005).

\bibitem{Almut}
        Y.-L.\ Lim, A.\ Beige, and L.C.\ Kwek,
Phys.\ Rev.\ Lett.\ {\bf 95}, 030505 (2005).

\bibitem{BenjComment}
	S.C.\ Benjamin,
	quant-ph/0504111.

\bibitem{SchlingeOld}
	D.\ Schlingemann and R.F.\ Werner, 
	Phys.\ Rev.\ A {\bf 65}, 012308 (2002).

\bibitem{TheStabil}
	Its stabilizer is given by the abelian group generated by
	the mutually commuting operators
	\begin{equation}
		K_a = \sigma^{(a)}_x
		\prod_{b \in N_a}
		\sigma^{(b)}_z,
	\end{equation}
	$N_a$ denoting the set of neighbors of $a$.
	
\bibitem{MaartenPhD}
	M.\ van den Nest, PhD thesis (KU Leuven, June 2005).	

\bibitem{SchlingeEquiv}
	D.\ Schlingemann, Quant.\ Inf.\ Comp.\
	{\bf 3}, 431 (2003).

\bibitem{OtherExceptions}
	In optical lattices, graph state preparation has been considered
	in Refs.\ \cite{Ox,Jiannis}. A promising sequential generation
	of multi-qubit matrix-product states -- and hence also 
	cluster states -- in time-bin encoding using
	optical cavities 
	has in turn been considered in Ref.\ \cite{Enrique}.

\bibitem{Ox}
	S.R.\ Clark, C.\ Moura Alves, and D.\
Jaksch, New J.\ Phys.\ {\bf 7}, 124 (2005).

\bibitem{Jiannis}
        A.\ Kay, J.K.\ Pachos, and C.S.\ Adams, quant-ph/0501166.

\bibitem{Enrique}
	C.\ Sch{\"o}n, E.\ Solano, F.\ Verstraete, J.I.\ Cirac, and
	M.M.\ Wolf, quant-ph/0501096.

\bibitem{BellLambda}
	http://www.bell-labs.com/news/1999/november/10/1.html

\bibitem{NL1}
        J.\ Eisert, K.\ Jacobs, P.\ Papadopoulos,
        and M.B.\ Plenio,
        Phys.\ Rev.\ A {\bf 62}, 052317 (2000).

\bibitem{NL2}
        D.\ Collins, N.\ Linden, and S.\ Popescu,
        Phys.\ Rev.\ A {\bf 64}, 032302 (2001).

\bibitem{NL3}
        J.I.\ Cirac, W.\ D{\"u}r, B.\ Kraus, and
        M.\ Lewenstein,
        Phys.\ Rev.\ Lett.\ {\bf 86}, 544 (2001).

\bibitem{Svore}
        K.M.\ Svore, B.M.\ Terhal, and D.P.\ DiVincenzo,
        quant-ph/0410047.

\bibitem{sta03}
	T.M.\ Stace, G.J.\ Milburn, and C.H.W.\ Barnes,
	 Phys.\ Rev.\ B {\bf 67}, 085317 (2003).

\bibitem{newBrowne}
	M.\ Varnava, D.E.\ Browne, and 
	T.\ Rudolph, quant-ph/0507036.

\bibitem{Remark}	
	It is an open problem whether to consider local Clifford
	unitaries is a restriction of generality when asking whether
	two graph states are LU-equivalent. In a large number of 
	cases for $n$ qubit graph states, 
	one can show that it indeed 
	suffices to take Clifford unitaries into account 
	\cite{MaartenLUEquivalent,Gross}.
		
\bibitem{MaartenLUEquivalent}
	M.\ van den Nest, J.\ Dehaene, and B.\ De
Moor, Phys.\ Rev.\ A {\bf 71}, 062323 (2005).


\bibitem{Gross}
        D.\ Gross, Diploma thesis (University of Potsdam,
        July 2005).

\bibitem{Perdrix}
	M.\ Mhalla and S.\ Perdrix, quant-ph/0412071.

\bibitem{MaartenEquivalent}
	M.\ van den Nest, J.\ Dehaene, and B.\ de Moor,
	Phys.\ Rev.\ A {\bf 70}, 034302 (2004).
		
\bibitem{Prehist}
An amusing analogy for this approach is provided by the method that  
prehistoric builders are supposed to have used to transport  
monolithic stones such as those forming stonehenge. The stone might  
be rolled on a set of tree-trunks -- as each tree -- 
trunk emerged from  
behind the stone, it would be quickly brought around the front `just  
in time' for the stone to roll on to it. Thus they would require only  
a few tree-trunks rather than paving the entire route with such  
rollers -- a tremendous saving in resources.

\bibitem{Elham}
	V.\ Danos, E.\ Kashefi, and P.\ Panangaden,
	quant-ph/0412135.

\bibitem{gar00}
	C.W.\ Gardiner and P.\ Zoller, {\it Quantum noise}
	(Springer, Heidelberg, 2000).

\bibitem{Knight}
	M.B.\ Plenio and P.L.\ Knight,
	Rev.\ Mod.\ Phys.\  {\bf 70}, 101 (1998).

\bibitem{Holevo}
	A.S.\ Holevo, {\it Statistical structure of quantum theory}
	(Springer, Heidelberg, 2001).
	
\bibitem{Gilchrist} A.\ Gilchrist, personal 
	communication.
	
\end{thebibliography}
\end{document}